\documentclass[aps,preprint,amsmath,amssymb]{revtex4-1}
\usepackage[dvipdfmx]{graphicx}% Include figure files
\usepackage{here}
\usepackage{txfonts}
\usepackage{color}

\begin{document}
\title{Magnetic field hysteresis of thermal conductivity in type-I and typy-II superconductors}
\author{Hiroto Arima$^1$\footnote{h-arima@aist.go.jp}}
\author{Miku Yoshida}
\author{Yoshikazu Mizuguchi$^{1}$}
\affiliation{$^1$Department of Physics, Tokyo Metropolitan University, Hachioji, Tokyo, 192-0397, Japan}
\begin{abstract}
This study investigated the magnetic field ($H$) dependence of thermal conductivity ($\kappa$) in type I (Pb) and type II (Nb) superconductors. The $\kappa$ of Nb displayed hysteresis, showing a local minimum during increasing $H$ process but not during decreasing $H$ process. Different magnetic field dependencies were observed in Pb with varying purity: $\kappa$ of 3N-Pb exhibited broad hysteresis, while that of 5N-Pb showed hysteresis similar to 3N-Nb.
\end{abstract}
\maketitle
In recent years, research endeavors have been undertaken to control thermal conductivity ($\kappa$) through the utilization of external fields, specifically electric and magnetic fields ($H$), without the need for mechanical moving\cite{Hirata2023,Ihlefeld2015,Kimling2013}. Superconductors are being studied as magneto-thermal switching (MTS) materials that operate at low temperatures, taking advantage of the difference in $\kappa$ between the normal and superconducting states because the Cooper pair does not transport heat\cite{DIPIRRO2014172,Yoshida_2023_Nb,Yoshida_2023_Pb}. Recently, Sn-Pb solders, which are almost phase-separated, have attracted attention as nonvolatile MTS materials maintaining a high $\kappa$ value at zero field even after experiencing $H$\cite{Arima_2023_solder}. Notably, the origin of nonvolatile MTS in Sn-Pb solders arises from the trapping of magnetic flux in the Sn region at zero field after experiencing $H$, thereby suppressing the superconducting bulkiness. Nonvolatile MTS caused by a similar mechanism has also been reported for MgB$_2$\cite{Arima_2023_MgB2}. To develop a deeper discussion of the nonvolatile MTS in superconductors, it is important to comprehensive research on the $H$ dependence of $\kappa$, including the decreasing $H$ process, in various superconductors. Although there are many reports of thermal conductivity measurements in magnetic fields because it provides valuable insights into the superconducting gap and electron correlation of superconductors\cite{Uher1990,Ausloos1999}, there are few studies that delve into the hysteresis of $\kappa$ in magnetic fields\cite{Peacor1991}. This study focused on pure metals as a fundamental research of the nonvolatile MTS in superconductors. In this study, we conducted thermal conductivity measurements of $H$ dependence including the decreasing $H$ process of type-II superconductor 99.9 \% purity of Nb (3N-Nb), type-I superconductors 99.9 \% and 99.999 \% purities of Pb (3N-Pb, 5N-Pb). 

The $H$ dependence of $\kappa$ for the 3N-Nb (polycrystalline, 3N purity, Nilaco) sheet with a thickness of 0.1 mm, 3N-Pb (polycrystalline, 3N purity, Nilaco), and 5N-Pb (polycrystalline, 5N purity, Nilaco) wires with a diameter of 0.5 mm, was measucite using a four-terminal configuration. These measurements were conducted with the thermal transport option (TTO) with the physical property measurement system (PPMS, Quantum Design).  The samples were mounted to ensure that the direction of the heat flow was antiparallel to the magnetic field. X-ray fluorescence (XRF) was conducted to assess the elemental composition of each sample. XRF results indicated the absence of any elements other than the primary elements in 3N-Nb and 5N-Pb. However, 3N-Pb exhibited the presence of Cu with a mass concentration of 0.07 \%.

Figure 1(a) illustrates the $H$ dependence of $\kappa$ in 3N-Nb at 2.5 K. As the magnetic field was applied, $\kappa$ maintains a constant value up to the lower critical field ($H_{\rm c1}$). Upon exceeding $H_{\rm c1}$, $\kappa$ decreases, reaching a local minimum at approximately 900 Oe. Beyond 900 Oe, $\kappa$ exhibits an increase. This upward trend in $\kappa$ continues up to the upper critical field ($H_{\rm c2}$), but beyond $H_{\rm c2}$, $\kappa$ shows $H$ independence. During the decreasing $H$ process, $\kappa$ remains constant until $H_{\rm c2}$. However, once $H$ falls below $H_{\rm c2}$, $\kappa$ shows a decrease. Notably, $\kappa$ does not show a local minimum during the decreasing $H$ process and dose not return to its initial value. The $H$ dependence of $\kappa$ in Nb is commonly known to exhibit the local minimum during the application $H$ process\cite{Noto_1969,Wasim_1969}. However, it seems to be a less-understood phenomenon that the $H$ dependence of $\kappa$ during the decreasing $H$ process differs from that of the application process. To explain the magnetic hysteresis of Nb, we present a schematic image of the $H$ dependence of $\kappa$ in Figures 1 (b) and (c). As is widely know, the total thermal conductivity ($\kappa_{\rm total}$) consists of two components: the electron contribution ($\kappa_{\rm e}$) and the phonon contribution ($\kappa_{\rm ph}$). As illustrated in Fig. 1(b), in the Meissner state, $\kappa_{\rm e}$ is negligible, and $\kappa_{\rm total}$ is almost dominated by $\kappa_{\rm ph}$. However, when the $H$ exceeds $H_{c1}$, magnetic flux penetrates into Nb, the normal conducting region expands and $\kappa_{\rm e}$ increases, while $\kappa_{\rm ph}$ decreases due to phonons being scattecite by the magnetic flux\cite{Lowell_1970}. Consequently, above a certain magnetic field, $\kappa_{\rm e}$ becomes dominant, and $\kappa_{\rm total}$ exhibits a local minimum. This suggests that the thermal conductivity of the superconducting state is sensitive to the magnetic flux, and the contributions of $\kappa_{\rm e}$ and $\kappa_{\rm ph}$ differ in the Meissner and mixed states. In real Type II superconductors, magnetic flux is trapped, meaning that once the magnetic field is applied, the material does not return to the Meissner state. As depicted in Fig. 1(c), the reason why $\kappa$ shows the magnetic hysteresis is believed to be because $\kappa_{\rm ph}$ does not return to the initial value due to the trapped magnetic flux.

Figure 2(a) illustrates the $H$ dependence of $\kappa$ in 3N-Pb at 2.5 K. During the application $H$, $\kappa$ does not exhibit a local minimum and sharply rises beyond the critical field. Considering that the type-I superconductor does not allow magnetic flux penetration below critical field, $\kappa_{\rm ph}$ is assumed to remain constant irrespective of $H$. Thus, $\kappa$ of 3N-Pb shows an increase with a sharp rise in $\kappa_{\rm e}$ from superconductivity to normal state transition, but $\kappa$ shows no local minimum. The distinct characteristic of 3N-Pb is the broad hysteresis of $\kappa$ during the decreasing $H$ process. As mentioned above, 3N-Pb includes a small quantity of Cu as an impurity. As shown in Fig. 2 (b), the value of $\kappa$ for 5N-Pb is more than 100 times larger than that for 3N-Pb, suggesting that the thermal conductivity is sensitive to impurities. Therefore, flux trapping at impurity sites and grain boundaries may occur during decreasing $H$ process in 3N-Pb. Figure 2(b) illustrates the $H$ dependence of $\kappa$ for 5N-Pb at 2.5 K. In magnetic fields exceeding 600 Oe, the thermal conductivity is sufficiently large, making measurement challenging. The figure displays solely the magnetic field range in which the measurement was conducted, although the magnetic field was applied up to 1000 Oe before the decreasing $H$ process measurement. The $H$ dependence of $\kappa$ for 5N-Pb resembles to that of 3N-Nb: 5N-Pb shows a local minimum of $\kappa $ at about 500 Oe during the application $H$ process, no local minimum was observed during decreasing $H$ process, and $\kappa$ does not return to the initial value. Following the discussion of the $H$ dependence of $\kappa$ in 3N-Nb, the origin of the local minimum in 5N-Pb may be attributed to the decrease in $\kappa_{\rm ph}$. However, since Pb is a type-I superconductor, the Meissner state should be maintained up to the critical field, therefore both $\kappa_{\rm ph}$ and $\kappa_{\rm e}$ would be constant regardless of the magnetic field. One possibility is that the assumption of maintaining the Meissner state was incorrect; 5N-Pb might have become an intermediate state where magnetic flux penetrated, and the magnetic flux scatterd phonons. However, since the wire-shaped 5N-Pb is carefully mounted to align with the cylindrical and magnetic field directions, the shape effect is considecite small. Moreover, even if the absence of the local minimum can be explained by an intermediate state, both the initial zero-field state and the zero-field state after experiencing $H$ must be Meissner states, which cannot explain why $\kappa $ at the zero field is different. In order to discuss this point in more detail, it will be necessary to clarify the temperature dependence of $\kappa$ in the magnetic field and the $H$ dependence of $\kappa$ at several temperatures in more detail.

In this study we measucite the $H$ dependence of $\kappa$ of the type-I superconductor Pb and the type-II superconductor Nb. Nb exhibits different $H$ dependences of $\kappa$ with and without local minimum during increasing and decreasing $H$ processes, attributed to the different $H$ dependence of $\kappa_{\rm ph}$ due to trapped magnetic flux. Pb results are different with sample purity. Less pure 3N-Pb shows broad hysteresis, potentially due to magnetic flux trapping by impurities. In 5N-Pb, despite its high purity, the local minimum and hysteresis were observed, but it is difficult to clarify the origin of these in this study. Previous research on MTS using superconductor has highlighted factors like the operational temperature range and switching ratio. However, when employing a superconductor as a MTS, it becomes important that $\kappa$ reverts to its initial value at zero field, alongside considering the switching speed. As this study shows, even for pure metals, $\kappa$ in the magnetic field depends on the sample quality and the magnetic field history, and this study suggests that it is also important to study the switching operation with the ON/OFF of the magnetic field.

\section*{acknowledgment}
This work was partly supported by JST-ERATO (JPMJER2201), TMU Research Project for Emergent Future Society, and Tokyo Government-Advanced Research (H31-1).

\section*{Author contributions}
H. Arima and M. Yoshida contributed equally to this work.

\bibliographystyle{jpsj}
\bibliography{Reference}

\begin{figure}[H]
\begin{center}
\includegraphics[scale=1]{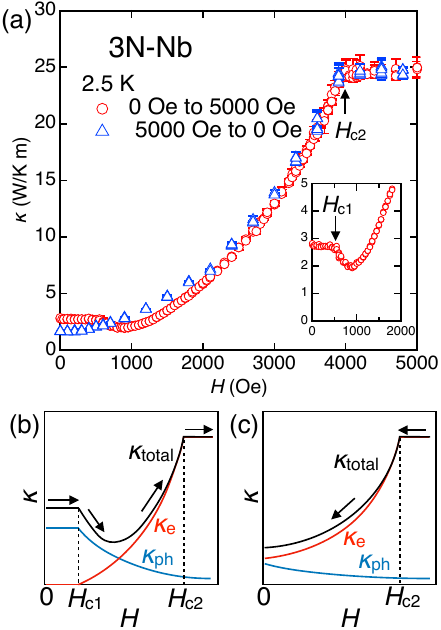}
\caption{(Color online) Magnetic field dependence of $\kappa$ for 3N-Nb at 2.5 K. The circular symbol signifies the magnetic field application process, while the triangular symbol signifies the demagnetization process. The insets show the magnetic field dependence of $\kappa$ near 1000 Oe.}
\label{Nb}
\end{center}
\end{figure}

\begin{figure}[H]
\begin{center}
\includegraphics[scale=1]{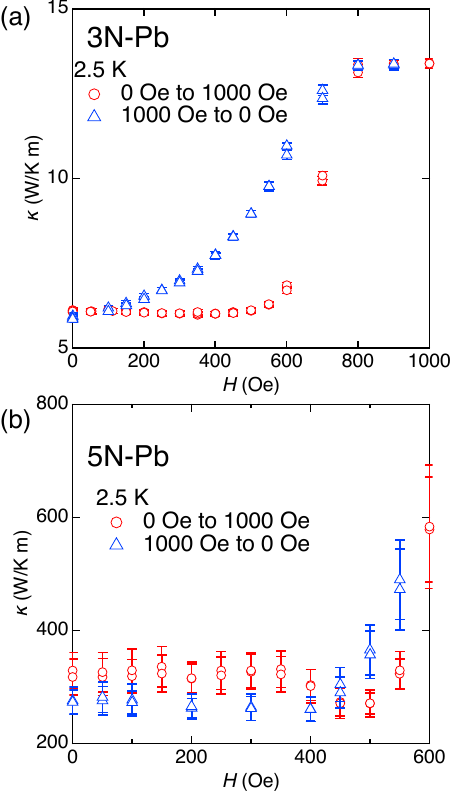}
\caption{(Color online) Magnetic field dependence of the $\kappa$ for (a) 3N-Pb and (b) 5N-Pb at 2.5 K. The circular symbol signifies the magnetic field application process, while the triangular symbol signifies the demagnetization process.}
\label{Pb}
\end{center}
\end{figure}

\end{document}